\documentclass{iopart}

\usepackage{iopams}
\usepackage{graphicx}

\begin{document}

\article[MACE]{Last Call for Predictions}{Mach Cones at central LHC
Collisions via MACE}

\author{
Bj\"orn B\"auchle$^{1,2}\footnote{Speaker}$,
Horst St\"ocker$^{1,3}$,
L\'aszl\'o P Csernai$^{2,4}$
}
\address{$^1$ Institut f\"ur theoretische Physik, Universit\"at Frankfurt,
Max-von-Laue-Stra\ss e 1, D-60438 Frankfurt am Main, Germany}
\address{$^2$ Section for Theoretical Physics, Departement of Physics,
University of Bergen, All\'egaten 55, 5007 Bergen, Norway}
\address{$^3$ Frankfurt Institute for Advanced Studies, Universit\"at Frankfurt,
Max-von-Laue-Stra\ss e 1, D-60438 Frankfurt am Main, Germany}
\address{$^4$ KFKI Research Institute for Particle and Nuclear Physics, P.O.
Box 49, 1525 Budapest, Hungary}

\eads{\mailto{baeuchle@th.physik.uni-frankfurt.de}}

\begin{abstract}

The shape of Mach Cones in central lead on lead collisions at
$\sqrt{s_{NN}} = 5.5$~TeV are calculated and discussed using MACE.

\end{abstract}

\submitto{\jpg}

\section{Introduction}

After the discovery of ``non-trivial parts'' in three-particle correlations
at RHIC \cite{Ulery:2007zb}, which are compatible with the existence of Mach
cones \cite{Satarov:2005mv}, it is interesting to see how the signal for
Mach cones will look like under the influence of a medium created at the
LHC in PbPb-Collisions. 

Mach cones caused by ultrarelativistic jets going in midrapidity will create
a double-peaked two-particle correlation function ${\rm d}N / {\rm d}(\Delta
\varphi)$. Those peaks are located at $\Delta \varphi = \pi \pm \cos^{-1}
c_{\rm S}$, where $c_{\rm S}$ is the speed of sound as obtained by the
equation of state. The model MACE (``Mach Cones Evolution'') has been
introduced to simulate the propagation of sound waves through a medium and
recognize and evaluate mach cones \cite{Baeuchle:2007qw}.

The medium is calculated without influence of a jet using the hydrodynamical
Particle-in-Cell-method (PIC) \cite{Clare:1986qj}. For the equation of
state, a massless ideal gas is assumed, so that $c_{\rm S} = 1/\sqrt{3}$ and
$\cos^{-1} c_{\rm S} = 0.96$.  The sound waves are propagated independently
of the propagation of the medium and without solving hydrodynamical
equations. Only the velocity field created by PIC is used.  To recognize
collective phenomena, the shape of the region affected by sound waves is
evaluated.

\section{Correlation functions}

\begin{figure}\begin{center} \includegraphics[width=.8\textwidth]{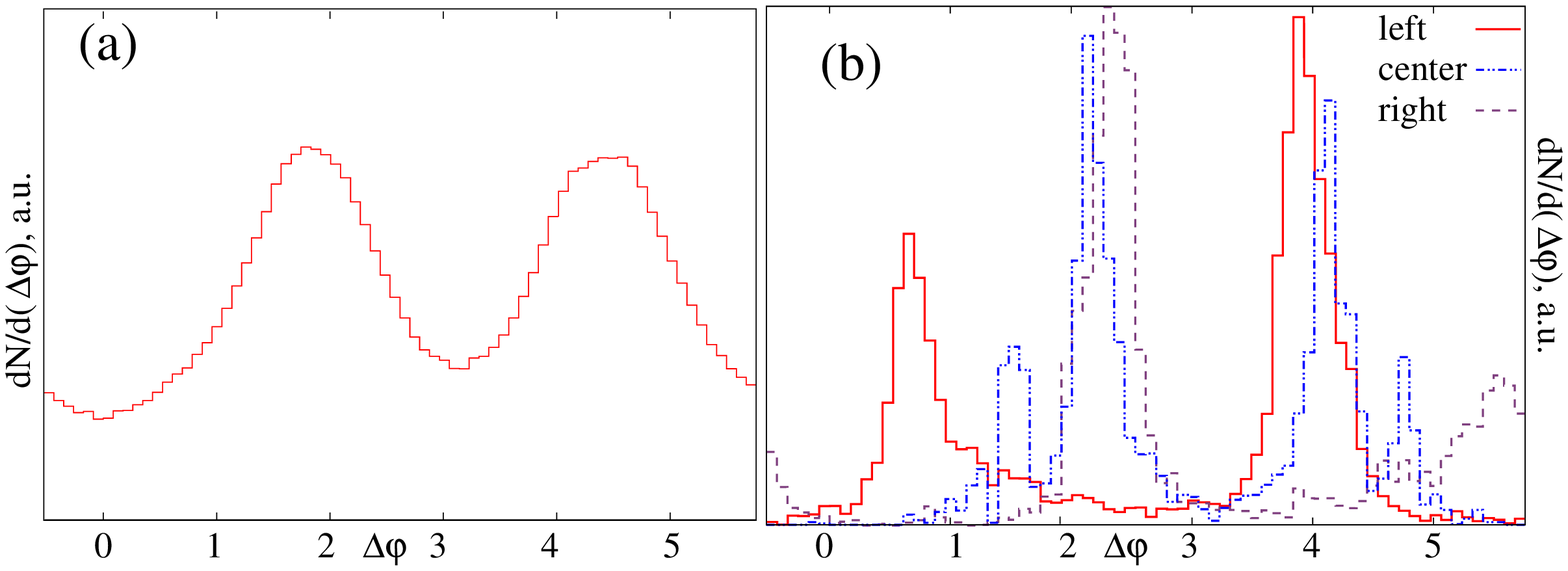}%
\caption{Two-particle-correlation function (away-side-part) for central
PbPb-Collisions at $\sqrt{s_{\rm NN}} = 5.5$~TeV. The peak created by the
forward jet is not calculated. (a): minimum jet bias (see text) with peaks
at $\Delta \varphi \approx \pi \pm 1.2$. (b): Midrapidity jets starting from
a position 70~\% on the way outside left and right of as well as in the
middle.} \label{fig} \end{center} \end{figure}

The correlation functions from the backward peak show a clear double-peaked
structure. The data for arbitrary jet origin and jet direction (minimum jet
bias) is shown in \fref{fig} (a). Here, the peaks are visible at $\Delta
\varphi \approx \pi \pm 1.2$. This corresponds to a speed of sound of
$c_{\rm S} \approx 0.36$. Note that the contributions from the forward
jet are not shown.
Deeper insight into different jet directions do not show a qualitatively
different picture.

\begin{figure}[b]\begin{center}
\includegraphics[width=.25\textwidth]{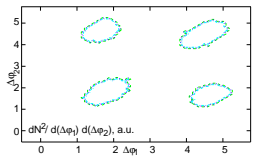}%
\caption{Three-particle-correlation function for the same data as in
\fref{fig}.} \label{3pc} \end{center} \end{figure}

Triggers on the origin of the jet, though, show the dependence of the
correlation function on the position where the jet was created (see
\fref{fig} (b)).
It shows that only the jet coming from the middle of the reaction results in
a symmetric correlation function with peaks at the mach angle $\Delta
\varphi = \pi \pm 0.96$. All other jets result in correlations that have
peaks at different angles, with the deviation getting bigger when going away
from the middle. Therefore, the speed of sound will always appear to be
smaller than it actually is.

\section{Conclusions}

If sound waves are produced from jet quenching in LHC-Collisions, the
two-particle correlation function will show the expected double-humped
structure in the backward region. The peaks will, though, be further apart
than $\delta(\Delta \varphi) = 2 \cos^{-1} c_{\rm S}$, thus alluding to a
speed of sound smaller than is actually present in the medium. 

The only case in which the true speed of sound can be measured is a
midrapidity jet that creates a symmetric correlation function.

\end{document}